\documentclass[showpacs,amsmath,amssymb,twocolumn,
superscriptaddress,aps,prl]{revtex4}
\usepackage{graphicx}

\begin{document}
\title{Giant Variations of Cooper-Pair Size in Nanoscale Superconductors}

\author{A. A. Shanenko}
\affiliation{Departement Fysica, Universiteit Antwerpen,
Groenenborgerlaan 171, B-2020 Antwerpen, Belgium}
\author{M. D. Croitoru}
\affiliation{Departement Fysica, Universiteit Antwerpen,
Groenenborgerlaan 171, B-2020 Antwerpen, Belgium}
\affiliation{University of Bayreuth, Institute of Theoretical
Physics, D-95440 Bayreuth, Germany}
\author{A. Vagov}
\affiliation{University of Bayreuth, Institute of
Theoretical Physics, D-95440 Bayreuth, Germany}
\author{F. M. Peeters}
\affiliation{Departement Fysica, Universiteit Antwerpen,
Groenenborgerlaan 171, B-2020 Antwerpen, Belgium}

\date{\today}

\begin{abstract}
The Cooper-pair size (i.e., the BCS coherence length) in
low-dimensional superconductors is dramatically modified by
quantum-size effects. In particular, for nanowires made of
conventional superconducting materials, we show that the coherence
length exhibits size-dependent drops by two-three orders of
magnitude and reaches values found in high-$T_c$ superconductors.
This phenomenon is surprisingly similar to the well-known BCS-BEC
crossover but with one important exception: it is driven by the
transverse quantization of the electron spectrum rather than by the
strength of the fermion-fermion interaction. Similar results can be
expected for other systems with the same structure of the
single-particle spectrum, e.g, for superfluid Fermi gases confined
in a quantum-wire or quantum-well geometry.
\end{abstract}

\maketitle

Superconductors of ultra-small dimensions possess unusual properties
not found in bulk materials. One of them is the {\itshape
quantum-size oscillations}, first discussed by Blatt and
Thompson~\cite{blatt}. In quasi-1D and -2D superconducting systems
(nanowires and nanofilms) quantization of the transverse electron
motion results in single-electron subbands, i.e., in multiple
quantum channels for the superconducting condensate. The proximity
of the lower edge of a single-electron subband to the Fermi surface
leads to a size-dependent enhancement of superconducting properties,
i.e., {\itshape the superconducting resonance}. In particular, such
resonances are expected to strongly influence the critical
temperature and critical magnetic field~(see, e.g.,
Refs.~\onlinecite{blatt,strong,sh,sh1}). Furthermore, they can lead
to a remarkable cascade structure of the superconductor-to-normal
transition~\cite{sh1} and can result in the appearance of a new type
of Andreev states induced by quantum confinement~\cite{sh2}.

For conventional materials, e.g., ${\rm Al}$, ${\rm Sn}$ or ${\rm
Pb}$, the superconducting gap is about $0.1-1.0\,{\rm
meV}$~\cite{degen} and, so, the inter-subband energy spacing
$\frac{\hbar^2}{2 m_e}\frac{\pi^2}{d^2}$~(with $d$ the confining
dimension) becomes of the same order or larger for $d \lesssim
20-40\,{\rm nm}$, where quantum-size oscillations of the
superconducting properties are expected to be significant. Several
recent experimental results on superconducting ${\rm Pb}$
nanofilms~\cite{guo,eom} and superconducting aluminum/tin
nanowires~\cite{arut} have been attributed to these quantum-size
effects (see Refs.~\onlinecite{guo,eom} and \onlinecite{sh},
respectively).

In the present Letter we report an unexpected phenomenon which is
due to quantum-size effects, i.e., giant variations of the
Cooper-pair size (i.e., the BCS coherence length $\xi_0$) in
low-dimensional superconductors. In all previous theoretical studies
of superconducting nanowires and nanofilms, e.g., modeling
phase-slip effects in nanowires (see, e.g., Ref.~\onlinecite{arut}),
one assumes that $\xi_0$ is specified by the same expressions as in
bulk. Contrary to this common expectation, our numerical
investigations of the Bogoliubov-de Gennes equations~\cite{degen}
for a clean superconducting quantum wire made of conventional
materials reveal that, depending on the wire width, the longitudinal
Cooper-pair size varies several orders of magnitude, from values of
a few micrometers, typical for conventional bulk superconductors, to
a few nanometers, that is usually found in hight-$T_c$
materials~\cite{high}. This phenomenon turns out to be very similar
to the BCS-BEC crossover in superfluid Fermi gases~\cite{bloch}.
However, in the present case a giant drop in the Cooper-pair size is
induced by quantum-size effects rather than by a change in the
strength of the fermion-fermion interaction. Our results are not
only relevant for superconducting quantum wires but are also
applicable to other systems with a similar single-particle spectrum,
e.g., to ultrathin superconducting metallic nanofilms and ultracold
Fermi gases confined in a quantum-wire or quantum-well geometry.

{\em Model and formalism}. -- We consider a superconducting
nanocylinder in the clean limit. For our numerical calculations we
take the material parameters of aluminum, the same as in
Refs.~\cite{sh,sh1,sh2}: $\hbar\omega_D =32.31\,{\rm meV}$; $gN(0) =
0.18$ (with $N(0)$ the bulk DOS); and $E_F = 0.9\,{\rm eV}$ is the
effective Fermi level in the parabolic band approximation (for more
details, see Ref.~\onlinecite{sh}). Two values for the wire diameter
are investigated below: $d=4.22\,{\rm nm}$, for which the wire is in
the resonance conditions, i.e., the bottom of one of the
single-electron subbands is close to the Fermi level; and
$d=4.35\,{\rm nm}$, when the wire is not influenced by a
superconducting resonance. We note that our conclusions are not
sensible to a particular choice of the parameters as long as $d
\lesssim 10-15\,{\rm nm}$; in the wires of larger diameters
quantum-size effects play less serious role.

Internal structure of Cooper pairs is described by $\Psi({\bf
x}_1,{\bf x_2})$~(the Cooper-pair wave function) which is
related to the anomalous Green's function as
\begin{equation}
\Psi({\bf x}_1,{\bf x}_2) = \imath \lim\limits_{t_1\to t_2+
0}{\cal F}({\bf x}_1t_1,{\bf x}_2t_2), \label{psi}
\end{equation}
where ${\cal F}({\bf x}_1t_1,{\bf x}_2t_2)=\frac{1}{\imath}
\langle {\cal T} \psi_{\uparrow}({\bf x}_1t_1)\psi_{\downarrow}
({\bf x}_2t_2)\rangle$ (for the spin-singlet pairing). In what
follows we use cylindrical coordinates ${\bf x} = (\rho,\varphi,
z)$. The diagonal part of Eq.~(\ref{psi-z}), i.e., at ${\bf x}_1
={\bf x}_2={\bf x}$, is directly related to the superconducting
order parameter $\Delta({\bf x})=g\Psi({\bf x},{\bf x})$, where
$g>0$ is the Gor'kov coupling constant. Rotational and
translational (along the wire) symmetries of the system are
reflected in $\Psi({\bf x}_1,{\bf x}_2)$ which depends on
$\phi = \varphi_1 -\varphi_2$ and $z=z_1-z_2$. In turn, the
order parameter is a function of $\rho$ only. Here, we are
interested in the structure of a Cooper pair along the wire
and, therefore, consider the quantity
\begin{equation}
\Psi(\rho,z)=\Psi(\rho,\varphi,z_1+z;\rho,\varphi,z_1).
\label{psi-z}
\end{equation}
The anomalous Green's function can be expressed in terms of
the eigenstates of the BdG equations (see, e.g., \cite{degen})
which, following the system symmetry, are specified by the radial
quantum number $j$, azimuthal quantum number $m$ and wavevector
$k$ of the quasi-free particle motion along the wire. This
defines 1D subbands labeled as $(j,m)$. For $T=0$ the wave
function in Eq.~(\ref{psi-z}) can be written as a sum over
such subbands as
\begin{equation}
\Psi(\rho,z) = \sum\limits_{jm}\Psi_{jm}(\rho,z), \label{psi-z1}
\end{equation}
with the subband contribution given by
\begin{align}
\Psi_{jm}(\rho,z)=\frac{1}{(2\pi)^2}\int\!\!{\rm d}k \;u_{jmk}
(\rho) v^{\ast}_{jmk}(\rho)\; e^{\imath k z}. \label{psi-z-jm}
\end{align}
Here $u_{jmk}(\rho)$ and $v_{jmk}(\rho)$ obey the BdG equations
written as
\begin{equation}
E_{jmk}\left(
\begin{array}{c}
u_{jmk}\\
v_{jmk}
\end{array}
\right)= \left(
\begin{array}{cc}
\hat{H}_{mk}        & \Delta(\rho)\\
\Delta^{\ast}(\rho) & -\hat{H}^{\ast}_{mk}
\end{array}\right)
\left(
\begin{array}{c}
u_{jmk}\\
v_{jmk}
\end{array}
\right), \label{BdG}
\end{equation}
with $E_{jmk}$ the quasiparticle energy and
$$
\hat{H}_{mk}=-
\frac{\hbar^2}{2m_e}\left[\frac{1}{\rho}\frac{\partial}{\partial
\rho}\rho\frac{\partial}{\partial\rho}-\frac{m^2}{\rho^2} -
k^2\right]-E_F,
$$
where $m_e$ is the electron band mass taken equal to the
free-electron mass. Solutions of Eq.~(\ref{BdG}) are set to be zero
at the wire boundary (quantum-confinement boundary conditions).

\begin{figure}[t]
\resizebox{0.92\columnwidth}{!}{\rotatebox{0}{\includegraphics%
{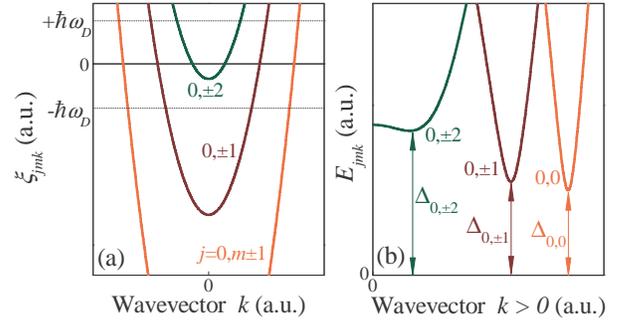}}}\caption{(Color online) (a) Single-electron energies
$\xi_{jmk}$ (measured from the Fermi level $E_F$) versus the
wavevector for the longitudinal motion $k$ in subbands
$(j,m)=(0,0),(0,\pm1)$ and $(j,\pm2)$. Horizontal dotted lines
denote the Debye window. (b) Quasiparticle energies $E_{jmk}$ as
function of $k$ for the same subbands.} \label{fig1}
\end{figure}

{\em Resonant subbands, qualitative picture}. -- The integral over
$k$ in Eq.~(\ref{psi-z-jm}) is restricted to the Debye window, i.e.,
$|\xi_{jmk}| < \hbar\omega_D$, where $\xi_{jmk}$ is the subband
single-particle dispersion $\xi_{jmk} = \hbar^2k^2/2m_e -\mu_{jm}$
with $\mu_{jm} = E_F-\varepsilon_{jm}$ the subband chemical
potential and $\varepsilon_{jm}$ the subband lower edge (bottom)
energy. Figure~\ref{fig1}(a) shows a sketch of the single-particle
energies for the subbands with $j=0$ and $m=0,\pm1,\pm2$. The
dotted horizontal lines in Fig.~\ref{fig1}(a) highlight the Debye
window that determines the upper $k^+_{jm}$ and lower $k^-_{jm}$
limits for $k$ in the integral in Eq.~(\ref{psi-z-jm}).

The bottoms of all single-electron subbands shift in energy with
changing diameter. A quantum-size superconducting resonance occurs
when the bottom of a subband comes into the Debye window, i.e., when
$|\mu_{jm}|< \hbar\omega_D$ and $k^-_{jm}=0$. In Fig.~\ref{fig1}(a)
subbands $(0,\pm2)$ satisfy this condition. Below they are referred
to as resonant subbands. Any subband generates a quantum channel for
the formation of the superconducting condensate. In a simplified
picture one can utilize Anderson's approximate solution of the BdG
equations (see, e.g., Ref.~\cite{sh1}), which assumes that the
spatial dependence of both $u_{jmk}(\rho)$ and $v_{jmk}(\rho)$ is
given by the radial single-electron wave function $\vartheta_{jm}
(\rho)$~(proportional to the Bessel function of the first kind).
This leads to $E_{jmk} =(\xi_{jmk}^2 + \Delta_{jm}^2)^{1/2}$, with
$\Delta_{jm}$ the subband energy gap as schematically shown in
Fig.~\ref{fig1}(b).

\begin{figure*}[t]
\resizebox{1.75\columnwidth}{!}{\rotatebox{0}{\includegraphics%
{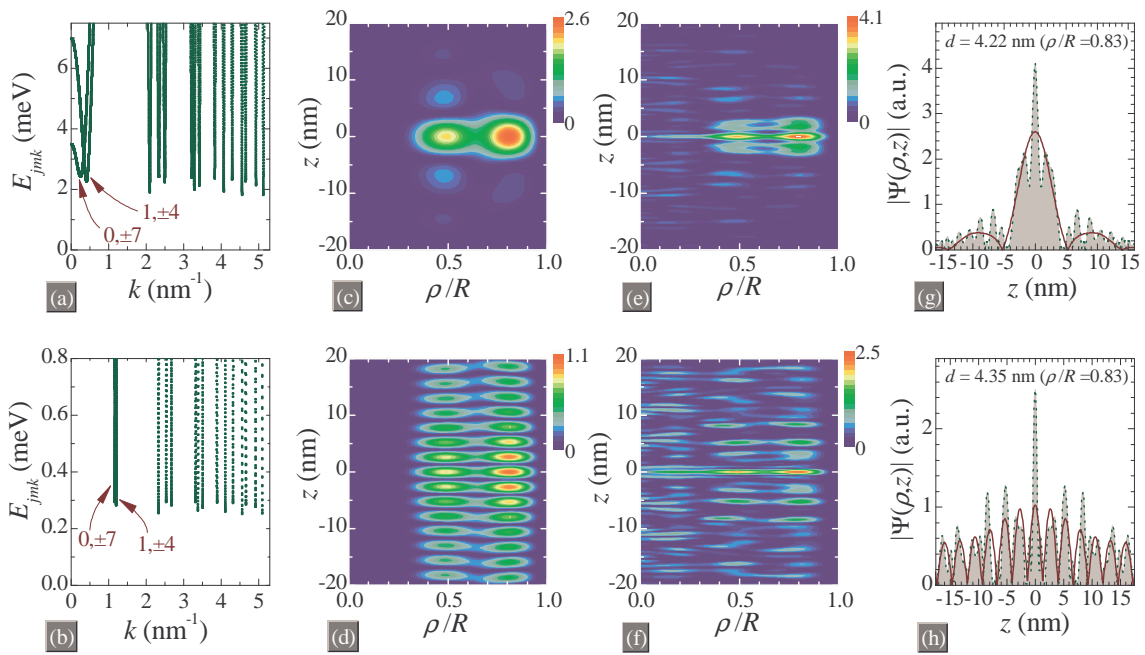}}}\caption{(Color online) The resonant wire, $d=4.22\,
{\rm nm}$: (a) the quasiparticle-energy dispersion $E_{jmk}$, the
resonant subbands are $(j,m)=(0,\pm 7)$ and $(1,\pm 4)$; (c) the
contour plot of $|\Psi(\rho,z)|$~(arbitrary units) when accounting
only for subbands $(0,\pm7)$ and $(1,\pm4)$; (e) the same but for
the total Cooper-pair wave function; (g) the longitudinal profile of
$|\Psi(\rho,z)|$ at $\rho/R=0.83$ with all relevant quantum channels
included (dotted curve) and when only contributions of $(0,\pm7)$
and $(1,\pm4)$ are taken~(solid red curve). Panels (b), (d), (e),
(f) display the same but for the non-resonant wire at $d=4.35\,{\rm
nm}$.} \label{fig2}
\end{figure*}

Within Anderson's approximation, Eq.~(\ref{psi-z-jm}) reduces to
\begin{equation}
\Psi_{jm}(\rho,z) = \frac{\vartheta_{jm}(\rho)^2}{(2\pi)^2}
\int\limits_{k^-_{jm}}^{k^+_{jm}}\!\!{\rm d} k
\frac{\Delta_{jm}\,\cos(kz)}{\sqrt{\xi^2_{jmk} +\Delta^2_{jm}}}.
\label{lcorr}
\end{equation}
In most cases the integration limits in Eq.~(\ref{lcorr}) can be
extended to infinity. This yields an exponentially decaying function
of $z$, and its characteristic decay length defines the subband
(channel) BCS coherence length
\begin{equation}
\xi_0^{(jm)} = \frac{\hbar}{\sqrt{m_e}}\left[\sqrt{\mu^2_{jm} +
\Delta_{jm}^2}-\mu_{jm}\right]^{-1/2}. \label{coh-length}
\end{equation}
As seen from Eq.~(\ref{coh-length}), $\xi_0^{(jm)}$ decreases when
$\mu_{jm}$ goes from positive to negative values. In the limit
$\mu_{jm}/\Delta_{jm} \gg 1$ we have the conventional result for the
BCS coherence length, which reads as $\xi_0^{(jm)} \approx \hbar
v_{jm}/\Delta_{jm}$, with $v_{jm} = \sqrt{2\mu_{jm} /m_e}$ the
subband Fermi wavevector. At resonance, i.e., for
$\mu_{jm}\rightarrow 0$, we find from Eq.~(\ref{coh-length}) a very
different expression $\xi_0^{(jm)} \approx \hbar/(m_e
\Delta_{jm})^{1/2}$. Finally, when $\mu_{jm} < 0$ and $|\mu_{jm}|
\gg \Delta_{jm}$, we obtain $\xi_0^{(jm)} \approx \hbar/(2m_e|
\mu_{jm}|)^{1/2}$, which decreases with increasing $|\mu_{jm}|$.
Thus, we have a drop in the BCS coherence length of the resonant
subband(s) and, at the same time, such a subband(s) provides a major
contribution to Eq.~(\ref{psi-z1}) due to a significantly enhanced
DOS in the Debye window.

\begin{figure}[t]
\resizebox{1.0\columnwidth}{!}{\rotatebox{0}{\includegraphics%
{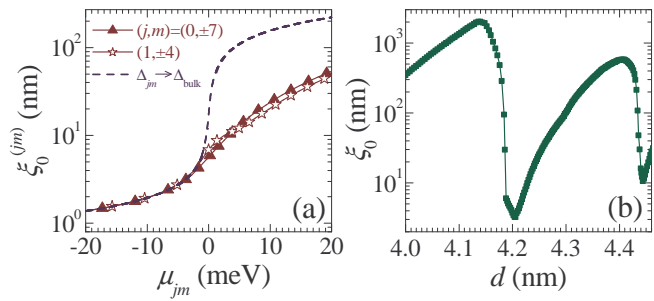}}}\caption{(Color online) BCS-BEC crossover induced by
quantum-size effects: (a) the subband longitudinal BCS coherence
length $\xi^{(jm)}_0$ versus $\mu_{jm}$ as numerically calculated
from the BdG equations for $(j,m)=(0,\pm7)$~(triangles) and
$(1,\pm4)$~(stars), the dashed curve represents
Eq.~(\ref{coh-length}) with $\Delta_{jm}$ replaced by $\Delta_{\rm
bulk}$; (b) the total longitudinal BCS coherence length $\xi_0$
versus the nanowire diameter.} \label{fig3}
\end{figure}

It is surprising that for $\mu_{jm} < 0$, Eq.~(\ref{coh-length}) is
reduced to the expression for the fermion-pair size at the BEC side
of the BCS-BEC crossover in superfluid Fermi gases~(see discussion
after Eq.~(140) in Ref.~\onlinecite{bloch}). However, in our case
$\mu_{jm}$ becomes negative due to quantum-size effects instead of a
change in the strength of the pair interaction (for numerical
details, see Fig.~\ref{fig3} and discussion below).

{\em Numerical solution}. -- A numerical self-consistent solution of
the BdG equations (\ref{BdG}) give the results shown in
Figs.~\ref{fig2} and \ref{fig3}. In Fig.~\ref{fig2}(a) the
quasiparticle energies $E_{jmk}$ are given versus $k$ for $d=4.22\,
{\rm nm}$. Here the bottoms of the two single-electron subbands with
$(j,m)=(0,\pm 7)$ and $(1,\pm 4)$ are in the Debye window and, so,
$v_{0,\pm7}$ and $v_{1,\pm4}$~(recall that $v_{jm} = \sqrt{2
\mu_{jm}/m_e}$) are extremely small. The quasiparticle spectrum for
the non-resonant wire with $d=4.35\,{\rm nm}$ is shown in
Fig.~\ref{fig2}(b). The corresponding single-electron spectrum has
no resonant subbands and $v_{0,\pm7}$ and $v_{1,\pm4}$ are larger by
an order of magnitude as compared to panel (a). Numerical results
for the resonant case exhibit a significant increase of
$\Delta_{0,\pm7}$ and $\Delta_{1,\pm4}$ and, in turn, lead to
enhanced superconducting gaps in the other quantum channels. When
the resonance decays (i.e., due to a change in $d$), all
$\Delta_{jm}$ are reduced and approach the bulk value $\Delta_{\rm
bulk}=0.25\,{\rm meV}$.

The contour plots in Figs.~\ref{fig2}(c) and (d), for the resonant
and non-resonant wires, respectively, display the absolute value of
$\Psi(\rho,z)$, as calculated from Eq.~(\ref{psi-z1}) but with the
summation restricted to subbands $(0,\pm 7)$ and $(1,\pm 4)$. Panels
(e)~[for $d=4.22\,{\rm nm}$] and (f)~[for $d=4.35\,{\rm nm}$] show
the total quantity $|\Psi(\rho,z)|$, where we summed over all
relevant subbands. For the resonant case, illustrated by panels (c)
and (e), the longitudinal distribution of electrons is well
localized, whereas an oscillating and weakly decaying dependence
appear in panels (d) and (f). This is also clearly seen from panels
(g) and (h) representing the longitudinal profile of $|\Psi(\rho,
z)|$ at $\rho/R=0.83$~[i.e., the maximum point of $\Psi(\rho,z=0)$]
for the resonant and non-resonant wires, respectively. Here the
dotted curve gives the total contribution of all subbands whereas
the solid line corresponds to a contribution from only
$(j,m)=(0,\pm7)$ and $(1,\pm4)$. Thus, the resonant subbands control
$\Psi(\rho,z)$ for the resonant wire, and the corresponding
longitudinal distribution of electrons in a Cooper pair is strongly
localized. At $d=4.35\,{\rm nm}$ single-electron subbands
$(j,m)=(0,\pm7)$ and $(1,\pm4)$ are shifted down as compared to
their positions at $d= 4.22\,{\rm nm}$. As a result, the resonance
disappears and the relative contribution of the states with $(j,m)=
(0,\pm7)$ and $(1,\pm4)$ to the superconducting order parameter
$\Delta (\rho)=\Psi(\rho,z=0)$ drops to $40\%$, see panel (h).
Nevertheless, the longitudinal decay of the correlation function
$\Psi(\rho,z)$ is still mainly determined by these states.

The longitudinal BCS coherence length is defined as the decay length
of $\Psi(\rho,z)$ in the $z$ direction and can be calculated through
a numerical fit. For both the partial and total wave functions, such
a fitting gives similar values: $\sim 1\,{\rm \mu m}$ for the non-%
resonant wire and $\approx 5\,{\rm nm}$ for the resonant case. The
latter value is almost three orders of magnitude less than the BCS
coherence length in bulk aluminum ($\approx 1.6\,{\rm \mu m}$) and
is comparable to the one in high-$T_c$ superconductors~\cite{high}.
Further insight can be obtained from Figs.~\ref{fig3}(a) and (b).
Panel (a) demonstrates numerical results for $\xi_0^{(jm)}$ as
function of $\mu_{jm}$ for $(j,m)=(0,\pm7)$~(triangles) and
$(1,\pm4)$~(stars). Notice that these data are in good agreement
with the analytical formula of Eq.~(\ref{coh-length}). When
substituting $\Delta_{\rm bulk}$ for $\Delta_{jm}$ in
Eq.~(\ref{coh-length}), we obtain the dashed curve approaching our
numerical results for $\mu_{jm} < 0$. In this case $\xi_0^{(jm)}
\approx \hbar/(2m_e|\mu_{jm}|)^{1/2}$ is not sensitive any more to
$\Delta_{jm}$. Finally, Fig.~\ref{fig3}(b) illustrates how the total
longitudinal BCS coherence length $\xi_0$ depends on $d$. Its value
is minimal at $d=4.22{\rm nm}$, and the difference between the
maximum and minimum is roughly two-three orders of magnitude. As
seen from Fig~\ref{fig3}(b), the next superconducting resonance
comes into play at $d=4.44\,{\rm nm}$. Notice that when increasing
$d$, quantum-size oscillations of $\xi_0$ are weakened and washed
out for $d \gtrsim 30-40\,{\rm nm}$.

{\em Conclusions}. -- We have demonstrated that the longitudinal
Cooper-pair size in a superconducting metallic nanowire undergoes
size-dependent giant drops driven by the transverse quantization of
the electron spectrum. As a result, the BCS coherence length of a
quantum wire made of conventional materials acquires values typical
for high-$T_c$ superconductors. A striking feature of this
phenomenon is that it is very similar to the BCS-BEC crossover found
in superfluid Fermi gases. However, there is a very important
difference: a giant drop of the longitudinal Cooper-pair size in our
quantum superconducting wire is a result of the transverse
quantization of the electron motion while the BCS-BEC crossover in
superfluid Fermi gases is realized by changing the inter-particle
interaction strength. Notice that the same qualitative behavior can
be found for different values of the material parameters and a sharp
Debye window is not essential. Therefore, such a phenomenon is a
generic feature that will be present in other low-dimensional
systems, where the condensate is formed via multiple quantum-size
channels, e.g., in superconducting nanofilms and ultracold Fermi
gases confined in a quantum-wire or quantum-well
geometry~\cite{bloch}.

\begin{acknowledgments}
This work was supported by the Flemish Science Foundation (FWO-Vl),
the Belgian Science Policy (IAP) and the ESF-network: INSTANS.
\end{acknowledgments}

\begin{acknowledgments}

\end{acknowledgments}

\end{document}